\documentclass[arx,a4paper,fleqn%
]{w-arx}
\usepackage{times,cite,w-thm}

\usepackage{mathptmx}
\usepackage{fontenc}
\usepackage{times}
\usepackage{amssymb}
\usepackage{graphicx}

\usepackage{amsmath}
\usepackage{amsfonts}
\usepackage{bbold}
\usepackage{inputenc}

\theoremstyle{plain}

\theoremstyle{definition}

\usepackage[]{graphicx}

\newcommand{\al}{\alpha}
\newcommand{\be}{\beta}

\usepackage{mathptmx}
\usepackage{fontenc}
\usepackage{times}
\usepackage{amssymb}
\usepackage{graphicx}

\usepackage{amsmath}
\usepackage{amsfonts}
\usepackage{bbold}
\usepackage{cite}
\usepackage{inputenc}
\def\5{\overline 5}

\newcommand{\diag}{\text{diag}}

\def\eq#1{{(\ref{#1})}}

\def\vev#1{\left\langle #1\right\rangle}

\def\Im{\mbox{Im}\,}

\def\hbar{\hspace{0pt}\raisebox{1pt}{$-$} \hspace{-7pt} h}

\newcommand{\beq}{\begin{equation}}
\newcommand{\eeq}{\end{equation}}
\newcommand{\bac}{\beq\begin{array}}
\newcommand{\eac}{\end{array}\eeq}
\newcommand{\ba}{\begin{array}}
\newcommand{\ea}{\end{array}}
\newcommand{\bea}{\begin{eqnarray}}
\newcommand{\eea}{\end{eqnarray}}
\begin{document}
\DOIsuffix{theDOIsuffix}
\pagespan{1}{}



\title[Lepton mixing induced by flavour symmetry and Leptogenesis constraints]{Lepton mixing induced by flavour symmetry and Leptogenesis constraints}


\author[I. de Medeiros Varzielas]{Ivo de Medeiros Varzielas\inst{1,2}%
  \footnote{Corresponding author\quad E-mail:~\textsf{ivo.de@udo.edu},
}}
\address[\inst{1}]{Fakult\"at f\"ur Physik, Technische Universit\"at Dortmund, D-44221 Dortmund, Germany}
\address[\inst{2}]{CFTP, Instituto Superior T\'ecnico}

\begin{abstract}
Flavour models may display a relation between the CP-violating asymmetry for leptogenesis and low-energy parameters. If the
flavour symmetry produces an exact mass independent lepton mixing scheme at leading order (with type I see-saw) the CP-violating asymmetry would vanish in the absence of corrections. We present a model displaying the link between deviations from the mixing scheme and leptogenesis.
\end{abstract}
\maketitle                   


This submission is based on \cite{ABdMM}, where a complete discussion with references is presented. Due to space constraints we focus here on the last part of the Corfu talk - proceedings covering the first part of the talk (also presented as a shorter talk in ICHEP) can be found in \cite{Ivo_ICHEP}.



The small neutrino masses can be
understood within the see-saw mechanism where
the Standard Model (SM) is extended by adding new heavy states - right-handed (RH)
neutrinos in type I.
The see-saw mechanism also contains the ingredients for leptogenesis. We discuss
only ``unflavoured'' leptogenesis, as in
the framework of flavour symmetry models the
heavy singlet neutrinos typically have masses above $10^{13}$~GeV and
for $T\gtrsim 10^{12}$ GeV lepton flavours are indistinguishable.
In the standard thermal leptogenesis scenario singlet neutrinos $N_{\al}$ are
produced by scattering processes after inflation. Subsequent out-of-equilibrium
decays of these heavy states generate a CP-violating asymmetry given
by
\begin{equation}
  \label{eq:cp-asymm}
  \epsilon_{N_\al} = \frac{1}{4v^2 \pi (m_D^{R\,\dagger} \;m_D^R)_{\al\al}}
  \sum_{\be\neq \al} \Im \left[\left((m_D^{R\,\dagger}
\;m_D^R)_{\be\al}\right)^2\right]
  f(z_\be)\,,
\end{equation}
where $z_\be=M_\be^2/M_\al^2$, $f(z_\be)$ is the loop function,
$m_D^R \equiv m_D V_R$ (i.e. in the basis where $M_R$ is diagonal).

We consider now a supersymmetric model based on $G_f=A_4\times Z_3\times Z_4$ \cite{Lin_Lepto} where relevant next to leading order (NLO) corrections appear only in the Dirac mass and can be parametrised with only 3 complex parameters. Neutrino masses arise only through type I see-saw so the results from \cite{ABdMM} apply.

$A_4$ is responsible for Tri-Bimaximal (TB) mixing, $Z_3\times Z_4$ generates the charged fermion hierarchy and avoids large mixing between the fields of the charged leptons and neutrino sectors.
The breaking sector includes scalar superfields $\varphi_T$, $\xi'$, $\varphi_S$, $\xi$ and $\zeta$. The assignments
are shown in table \ref{tab2}.

\begin{table}[h!]
\begin{center}
\begin{tabular}{|c|ccccccc|ccccc|}
\hline
&&&&&&&&&&&&\\[-3mm]
& $L$ & $e^c$ & $\mu^c$ & $\tau^c$ & $N^c$ & $H^u$ & $H^d$ & $\varphi_T$ & $\xi'$ & $\varphi_S$ & $\xi$ & $\zeta$ \\[0.3 mm]
\hline
&&&&&&&&&&&&\\[-3mm]
$A_4$ & $3$ & $1$ & $1$ & $1$ & $3$ & $1$ & $1$ & $3$ & $1'$ & $3$ & $1$ & $1$ \\[0.3 mm]
$Z_3$ & $1$ & $1$ & $1$ & $1$ & $\omega$ & $1$ & $1$ & $1$ & $1$ & $\omega$ &  $\omega$ &  $\omega^2$ \\[0.3 mm]
$Z_4$ & $1$ & $-i$ & $-1$ & $1$ & $1$ & $1$ & $-i$ & $i$ & $i$ & $1$ & $1$ & $1$ \\[0.3 mm]
\hline
\end{tabular}
\caption{Matter and scalar content of the model and their transformation properties under $G_f$ \cite{Lin_Lepto}.}
\label{tab2}
\end{center}
\end{table}
The LO Yukawa superpotential is expressed as an expansions in the cut-off of the theory $\Lambda$:
\bea
&&\ba{rcl}
\mathcal{W}_\ell&=&\dfrac{1}{\Lambda} y_\tau \left(L\varphi_T\right)\tau^cH^d+ \dfrac{1}{\Lambda^2} y_\mu^{(1)} \left(L\varphi_T\right)''\xi'\mu^cH^d+\dfrac{1}{\Lambda^2} y_\mu^{(2)} \left(L\varphi_T\varphi_T\right)\mu^cH^d+ \\ [3 mm]
&+&\dfrac{1}{\Lambda^3} y_e^{(1)} \left(L\varphi_T\right)'\left(\xi'\right)^2e^cH^d+\dfrac{1}{\Lambda^3} y_e^{(2)} \left(L\varphi_T\varphi_T\right)''\xi'e^cH^d+ \dfrac{1}{\Lambda^3} y_e^{(3)} \left(L\varphi_T\varphi_T\varphi_T\right)e^cH^d\;, \\
\ea\\
&&\hspace{3mm}\mathcal{W}_\nu=-\dfrac{1}{\Lambda} y \left(LN^c\right)\zeta H^u+ x_a\left(N^cN^c\right)\xi+ x_b\left(N^cN^c\varphi_S\right)\;,
\eea
where $(\ldots)$, $(\ldots)'$ and $(\ldots)''$ stand for the contraction in the representations $1$, $1'$ and $1''$ of $A_4$, respectively.

The flavon superfields acquire the following VEVs ($v_T$, $u'$, $v_S$, $u$ and $w$ are small parameters):
\beq \ba{lllll}
\vev{\varphi_T}= \left(
                \begin{array}{c}
                  0 \\
                  v_T \\
                  0 \\
                \end{array}
              \right)\;, &
\vev{\xi'}= u'\;, &
\vev{\varphi_S}=\left(
                \begin{array}{c}
                    v_S \\
                    v_S \\
                    v_S \\
                  \end{array}
                \right)\;,&
\vev{\xi}= u\;, &
\vev{\zeta}= w\;,
\ea
\eeq
and these VEVs can be aligned naturally from the scalar potential and lead to TB mixing \cite{Lin_Lepto}. The symmetry prevents deviations from these VEVs at NLO and allows the order of magnitude relations between parameters $v_T \sim u'$ and $v_S \sim u \sim w$, assuming at most a mild hierarchy among the two sets.

%

The neutrino mass matrix gets contributions from the type I see-saw. We have:
\bea
\label{neutLO}
m_D=\left(
        \begin{array}{ccc}
        1 & 0 & 0 \\
        0 & 0 & 1  \\
        0 & 1 & 0
        \end{array}  \right)\dfrac{y\,w\,v^u}{\Lambda}\,,
  &\quad&
  M_R= \left(
        \begin{array}{ccc}
        b+2d & -d & -d \\
        -d & 2d & b-d \\
        -d & b-d & 2d
        \end{array} \right)u\,,
\eea
with $v^u=\vev{H^u}$, $b\equiv2x_a$ and $d\equiv2x_bv_S/u$.
$M_R$ and $m_\nu$ are diagonalised by the TB mixing matrix $U_{TB}$, giving as eigenvalues $M_1=|b+3d|$, $M_2=|b|$, $M_3=|b-3d|$ and $m_i=(y\,w\,v^u)^2/(\Lambda^2M_i)$.
To estimate $\epsilon_{N_\alpha}$, we write the Dirac mass matrix in the basis of diagonal and real RH neutrinos:
\beq
m_D^R=m_DU_{TB}D'\,,
\eeq
where $D'=\diag(e^{i\phi_1/2},\,e^{i\phi_2/2},\,e^{i\phi_3/2})$ and $\phi_\al$ are the phases of $b+3d$, $b$, $b-3d$ respectively (eigenvalues of $M_R$).
$m_D^{R\dag} m_D^R$ is diagonal and thus $\epsilon_{N_\alpha}=0$, in agreement with the model-independent proof in \cite{ABdMM}.
A non-vanishing asymmetry can be obtained at NLO. In this
model $Z_3\times Z_4$ only admits NLO corrections to the Dirac terms, and the ones that can not be reabsorbed in a redefinition of the LO parameters are:
\beq
-\mathcal{W}^{NLO}_\nu=\dfrac{1}{\Lambda} y_1 \left(LN^c\right)'\left(\varphi_S\varphi_S\right)''H^u+ \dfrac{1}{\Lambda} y_2 \left(LN^c\right)''\left(\varphi_S\varphi_S\right)'H^u+\dfrac{1}{\Lambda} y_3 \left(\left(LN^c\right)_A\varphi_S\right)\xi H^u\,,
\eeq
where $(\ldots)_A$ refers to the asymmetric contraction of triplets. The deviations to $m_D$ can be written as
\beq
\label{md1}
m_D^{(1)}=
\left(
  \begin{array}{ccc}
    0 & y_1+y_3 & y_2-y_3 \\
    y_1-y_3 & y_2 & y_3 \\
    y_2+y_3 & -y_3 & y_1 \\
  \end{array}
\right)v^u\dfrac{v_S^2}{\Lambda^2}\,,
\eeq
where $y_3$ accounts for the ratio $u/v_S$.
Including \eq{md1}, the TB mixing receives small perturbations according to $U_\nu=U_{TB}\delta U$, where only the element $(\delta U)_{13}$ is relevant. Parametrising this term as:
\beq
(\delta U)_{13}=\sqrt{\dfrac{3}{2}}\sin\theta_{13}e^{i \delta }\sim\mathcal{O}\left(\dfrac{v_S}{\Lambda}\right)\,,
\eeq
where $\delta$ is the CP-violating Dirac phase in the standard parametrisation of the mixing matrix, we get
\beq
\label{expang}
\sin^2\theta_{23}=\dfrac{1}{2}(1+\sqrt2\cos \delta \sin\theta_{13}) \qquad\qquad\sin^2\theta_{12}=\dfrac{1}{3}(1+\sin^2\theta_{13})\,.
\eeq
We estimate NNLO perturbations of order $\sin^2\theta_{13}$ and thus $\sin^2\theta_{12}$ will receive non-negligible corrections.
We can impose an upper bound on $v_S/\Lambda$ by requiring that the correction to the TB value of $\sin^2\theta_{12}$ does not take it outside the experimental $3\sigma$ range: the maximal allowed deviation from the TB value is $0.05$ and from there we impose the bound $v_S/\Lambda<\mathcal{O}(0.23)$.
We consider now $m_D^{R\prime}$ (the NLO Dirac neutrino mass matrix in the basis of diagonal and real RH neutrinos). We can write:
\beq
\label{expmDRp}
m_D^{R\prime}=m_D^R+m_D^{(1)}U_{TB}D'\,,
\eeq
and calculate the relevant product for leptogenesis, $m_D^{R\prime\dag}m_D^{R\prime}$, keeping only the first terms in the expansion in the small parameter $v_S^2/\Lambda^2$:
\beq
\label{md2}
m_D^{R\prime\dag}m_D^{R\prime}=m_D^{R\dag}m_D^R+ \left(D^*U_{TB}^Tm_D^{(1)\dag}m_D^R+h.c.\right)\,,
\eeq
where in the second term the only off-diagonal entries are the $13$ and $31$ ones. In the case of inverted hierarchy for the effective neutrinos, the lightest RH neutrino is $N_2$. The summation in the numerator of eq. (\ref{eq:cp-asymm}) does not contain the term $13$ and therefore $\epsilon_{N_2}$ is vanishing also at NLO. This, however, does not mean leptogenesis can not be realized in this case. Since there is only a mild hierarchy between $N_2$, $N_1$ and $N_3$ and neither $\epsilon_{N_1}$ nor $\epsilon_{N_3}$ vanish, leptogenesis will proceed through $N_{1,3}$ dynamics. In the normal hierarchy (NH), the RH neutrino mass spectrum is $M_{N_3}<M_{N_2}<M_{N_1}$. There is a mild hierarchy between $N_3$ and $N_2$ while the hierarchy between $N_3$ and $N_1$ is large (around a factor 9). Consequently, the lepton asymmetry generated in $N_1$ decays will be, in general, erased by the lepton number violating interactions of $N_3$. Only $N_3$ dynamics becomes relevant for the generation of a lepton asymmetry in this case. Note that if the hierarchy between $N_1$ and $N_3$ decreases (as could be in the case of a quasi-degenerate spectrum), so it becomes mild, $N_1$ dynamics should be taken into account. Henceforth, for simplicity, we will consider only the NH case for which, according to eq. (\ref{eq:cp-asymm}), the CP-violating parameter $\epsilon_{N_3}$ can be written as
\beq
\label{expep}
\epsilon_{N_3}=\dfrac{1}{8\pi}\dfrac{1}{v_u^2\left(m_D^{R\,\dagger} \;m_D^R\right)_{11}}\mathbb{I}\mbox{m}\left[\left(\left(m_D^{R\,\dagger} \;m_D^R\right)_{13}\right)^2\right]f\left(\dfrac{M_1^2}{M_3^3}\right)\,.
\eeq

In the following figures we show a series of scatter plots related to the predictions of the model and the connections among low-energy observables and $\epsilon_{N_3}$. The points correspond only to the NH neutrino spectrum (in which we take $v_{S}/\Lambda\sim w/\Lambda=0.007\div0.23$, $\tan\beta=2\div50$ and we treat $y$, $y_1$, $y_2$ and $y_3$ as random numbers with modulus between $0.1$ and $2$).

\begin{figure}[ht!]
  \centering
\includegraphics[width=3.8cm]{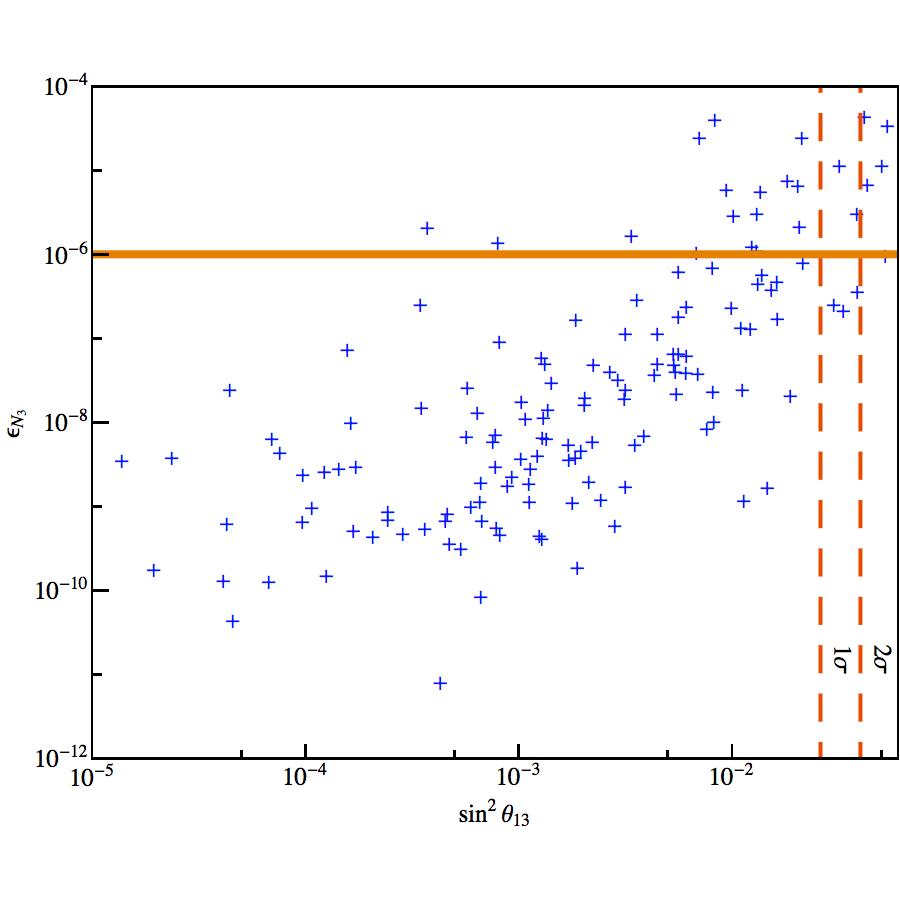}
\includegraphics[width=3.8cm]{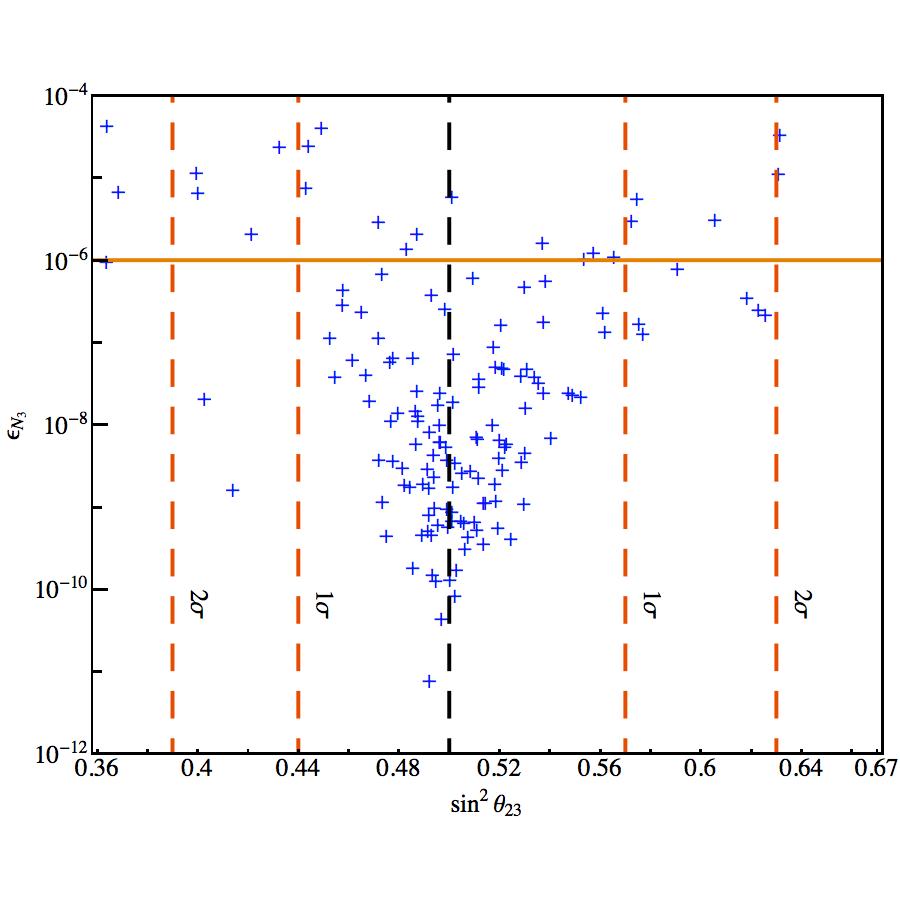}
\includegraphics[width=3.8cm]{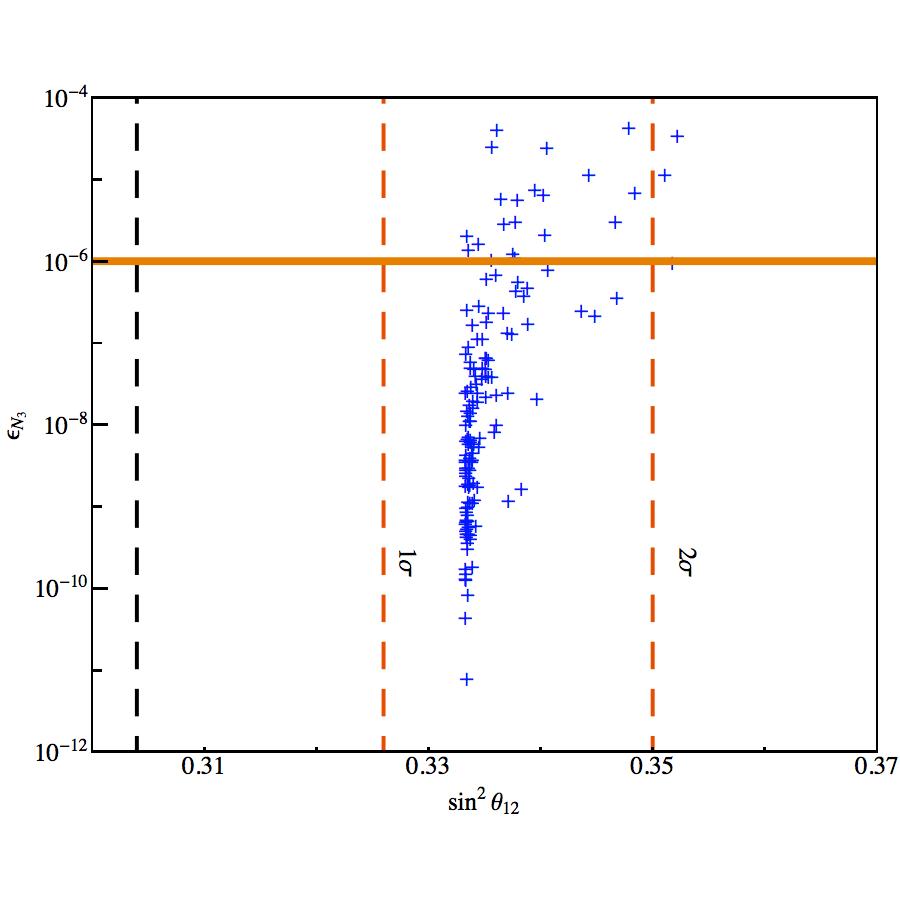} \caption{Correlation between $\epsilon_{N_3}$ and $\sin^2\theta_{13}$, $\sin^2\theta_{23}$, $\sin^2\theta_{12}$. The horizontal line marks $\epsilon_{N_3}\sim 10^{-6}$, the vertical lines correspond to the central values and bounds at $1$ and $2~\sigma$ level of $\sin^2\theta_{13, 23, 12}$.}
 \label{fig:eN-13-23}
\end{figure}

In figure~\ref{fig:eN-13-23} we plot $\epsilon_{N_3}$ and $\sin^2\theta_{13, 23, 12}$. As expected by comparing \eq{expep} with \eq{expang}, $\epsilon_{N_3}$ is correlated to all low-energy mixing angles.
\begin{figure}[ht!]
 \centering
\includegraphics[width=3.8cm]{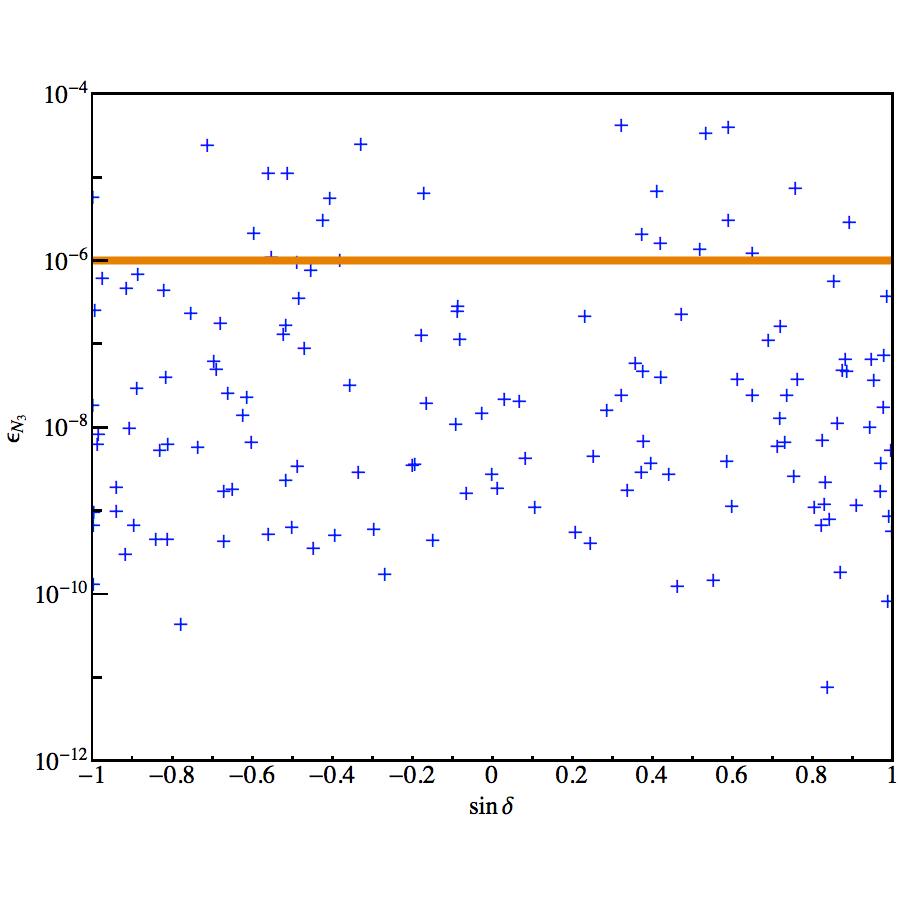}
\includegraphics[width=3.8cm]{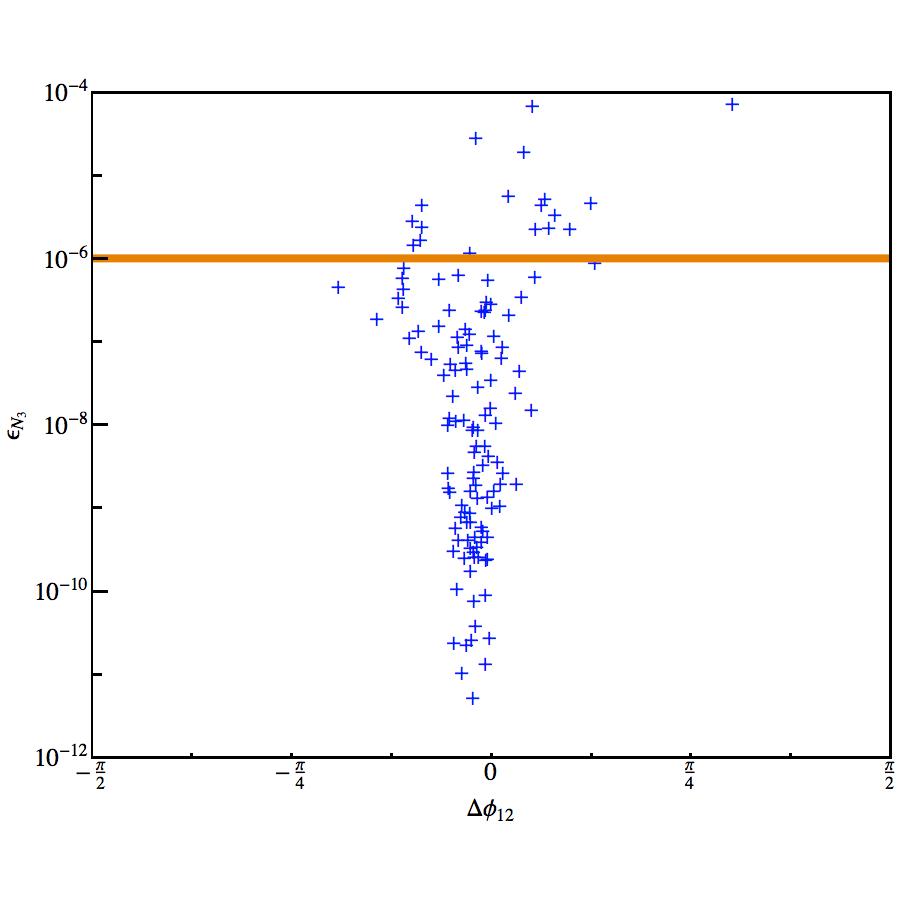}
\caption{The high energy $\epsilon_{N_3}$ versus low energy $\delta$ and $\Delta \phi_{12}$. The horizontal line marks to $\epsilon_{N_3}\sim 10^{-6}$.}
 \label{fig:del-lep}
\end{figure}
In figure~\ref{fig:del-lep} we plot $\epsilon_{N_3}$ and the low-energy CP-Dirac and Majorana  phases. It shows  that  the low energy  CP-Dirac phase  $\delta$ is not correlated to $\epsilon_{N_3}$. This result is not surprising: the phases that enter in $\epsilon_{N_3}$ are related to the phases  present in $M_R$, while $\delta$ arises by the  phases that appear in what we defined as $m_D^{(1)}$ in \eq{expmDRp}. In contrast the difference between the Majorana phases $\phi_1$ and $\phi_2$ ($\Delta \phi_{12}$) presents a correlation with
$\epsilon_{N_3}$: at LO the phenomenological analysis of this model shows that the NH spectrum can be reproduced only if  $\Delta \phi^0_{12} $ is small. At LO, $\Delta\phi^0_{12}$ coincides with the corresponding Majorana phase difference  $\Delta \phi_{12}^R$  of the right handed neutrinos. By perturbing the neutrino Dirac mass matrix we introduce new arbitrary phases that can vary in all the interval $(0,2 \pi)$. However the NLO contributions are responsible for deviating the lepton mixing angles away from TB values and also for slightly modifying the neutrino spectrum, in the range allowed by the data fit. This means that in general at NLO
\beq
m_i \sim m_i^0 +\delta m_i\,,
\eeq
where $\delta m_i$ are complex parameters and $m_i^0$  the neutrino mass eigenvalues at LO. Requiring now that $\Delta m^2_{12}$ is still in the range indicated for $\Delta m^2_{sol}$ we have that $\delta m\sim |\delta m_{1,2}| \sim 10^{-3}$ eV for  $|m_1^0|,|m_2^0|\sim \mathcal{O}(\sqrt{\Delta m^2_{sol}})$. A straight computation shows then that the Majorana phase $\Delta \phi_{12}$ satisfies
\beq
\tan  \Delta \phi_{12}\sim  \tan  \Delta \phi^0_{12}+ \alpha \frac{\delta m}{ \mathcal{O}(\sqrt{\Delta m^2_{sol}})}\,,
\eeq
where $\Delta \phi^0_{12}$ is the LO phase difference and $\alpha\in (0,1)$ a parameter that takes into account that the $\delta m_{1,2}$  phases run into the interval $(0, 2 \pi)$. We can estimate the maximal deviation of $\Delta\phi_{12}$ by its LO value getting
\beq
\Delta\phi_{12}- \Delta\phi^0_{12} \sim \frac{\pi}{10}\,.
\eeq
Notice that the left panel of fig.~\ref{fig:del-lep} shows that the majority of all the points are indeed inside the interval $(-\pi/8,\pi/8)$ in perfect agreement with our  analytical results  for a small LO  $\Delta\phi^0_{12}\leq0.1$.

In conclusion, in the model considered it is possible to obtain correlations between low-energy observables and high-energy CP-violation, but it confirms that in general no correlation is present between high and low-energy CP-violating parameters.
There is correlation between deviations from TB angles (e.g. non-vanishing value of $\theta_{13}$) and the
value of the CP asymmetry parameter. We note that to have viable leptogenesis the model's allowed parameter space is rather constrained.

\begin{acknowledgement}
The work of IdMV was supported by FCT under the grant
SFRH/BPD/35919/2007 and
through the projects POCI/81919/2007, CERN/FP/83503/2008 and CFTP-FCT
UNIT 777 which are partially funded through POCTI (FEDER) and by the
Marie Curie RTN MRTN-CT-2006-035505.
\end{acknowledgement}

\end{document}